\documentclass[aps,nopacs,pra,amssymb,amsmath,twocolumn]{revtex4-2}
\usepackage{graphicx,subfigure}
\usepackage{float}
\usepackage{amssymb}
\usepackage{amsmath}
\usepackage{xcolor} 
\usepackage{multirow}
\usepackage{hyperref}
\usepackage{booktabs}
\usepackage{textcomp}
\usepackage{cleveref}
\usepackage{verbatim}
\usepackage{bbm}
\usepackage{cancel}
\usepackage{comment}
\usepackage{braket}
\usepackage{kotex}
\usepackage{physics}
\usepackage{cleveref}
\usepackage{qcircuit}

\begin{document}
\title{Rydberg wire gates for universal quantum computation}

\author{Seokho Jeong$^{1}$, Xiao-Feng Shi$^{2}$, Minhyuk Kim$^{1}$, and Jaewook Ahn$^{1}$}
\email{jwahn@kaist.ac.kr}
\address{$^{1}$Department of Physics, KAIST, Daejeon 34141, Republic of Korea \\
$^{2}$School of Physics, Xidian University, Xi’an 710071, China}
\date{\today}

\begin{abstract} \noindent
Rydberg atom arrays offer flexible geometries of strongly-interacting neutral atoms, which are useful for many quantum applications such as quantum simulation and quantum computation. Here we consider a gate-based quantum computing scheme for a Rydberg-atom array. We utilize auxiliary atoms which are used as a quantum wire to mediate controllable interactions among data-qubit atoms. We construct universal quantum gates for the data atoms, by using single-atom addressing operations. Standard one-, two-, and multi-qubit solutions are explicitly obtained as respective sequences of pulsed operations acting on individual data and wire atoms. A detailed resource estimate is provided for an experimental implementation of this scheme in a Rydberg quantum simulator.
\end{abstract}

\maketitle

\section{Introduction} \noindent
Quantum computing is being actively studied as a means to revolutionize the humankind's computational capability beyond the limits of digital computers~\cite{Feynman1982,Nielsen}. Quantum computing hardware are physical two-level systems, which we refer to as qubits hereafter, and quantum computation performs operations of universal quantum gates on them. Gate-based quantum computations have been demonstrated in many physical systems, including linear optics~\cite{O'brien2007,Kok2007}, circuit quantum electrodynamics of superconductor~\cite{Arute2019,Krantz2019,Blais2021}, trapped ions~\cite{Kielpinski2002,Schindler2013,Pogorelov2021}, defects in solid-state materials~\cite{Sar2012,Childress2013}, and neutral atoms~\cite{Weiss2017,Henriet2020}. 

Neutral atoms have been considered for gate-based quantum computations using interactions between Rydberg atoms~\cite{Saffman2010,Saffman2016}. The advantages of using Rydberg atoms are strong dipole-dipole interactions that can be switched on and off by fast laser excitation, large-scale atom arrays that can be prepared with almost any desired geometries and topologies, and stable ground hyperfine states that can be used for long-term quantum information. Quantum gates using Rydberg atoms can utilize distance-dependent interactions~\cite{Jo2020} or the Rydberg blockade effect which prohibits adjacent atoms from being excited to a Rydberg state~\cite{Gaetan2009,Urban2009}. There are many Rydberg-atom schemes for quantum gates and entanglements~\cite{Jaksch2000,Shi2018,Shi2020,Saffman2020} and experimental demonstrations~\cite{Isenhower2010,Levine2018,Levine2019,Graham2019,Madjarov2020,Graham2022,Schine2021}. The fidelities of the recent demonstrations were recorded 0.97 in alkali atom system~\cite{Levine2019} and 0.99 in alkaline-earth atomic system~\cite{Madjarov2020}. Many of these previous studies are based on coding quantum information in the stable states, which are usually the hyperfine-Zeeman sub-states. For large-scale universal quantum computation, the required fidelity for each quantum operation in the circuit model of quantum computation should be high, which motivated the use of stable states for coding quantum information. However, our analyses to be discussed below show that gate fidelity can exceed 98\% for key elements in quantum computation, such as controlled-$\bf Z$ ($
\bf CZ$) gates, without resorting to the ground sub-levels. 

In this paper, we construct universal quantum gates using single-atom addressings in a Rydberg-atom array. We choose a Rydberg state to be one of the two qubit states for a data qubit and use a cluster of data and auxiliary qubits in a Rydberg atom array, in which the ancillary atoms between data qubits mediate interactions between the data atoms. The advantage of this setup comes in twofold. First, the distance between data atoms can be large, for which analyses shown later with practical and currently available resources tell us that a ${\bf CZ}$ gate between two atoms separated about $19~\mu$m can be created with a fidelity over 98\% within a duration $2\pi/\Omega$, where $\Omega$ is the Rydberg Rabi frequency. Second, the gates are all realized with fast laser excitation of ground-Rydberg transitions, so that the quantum circuit for a certain computational task~(including  digital quantum simulation) can be carried out fast. 

In the rest of the paper, we first outline the main idea of the quantum-wire gates based on the Rydberg interaction and single-atom addressing in Sec.~\ref{single-atom}, and then construct single and two-qubit gates in Secs.~\ref{one-qubit} and \ref{two-qubit}. We then discuss general two-qubit state generation and multi-qubit gates in Secs.~\ref{arbitrary} and \ref{multi}. Experimental implementations, gate performances, and alternative schemes are discussed in Sec.~\ref{discussion}.

\section{Single-atom addressing in a Rydberg-atom system}\label{single-atom} 
\begin{figure}[h]
\centering
\includegraphics [width=0.48\textwidth]{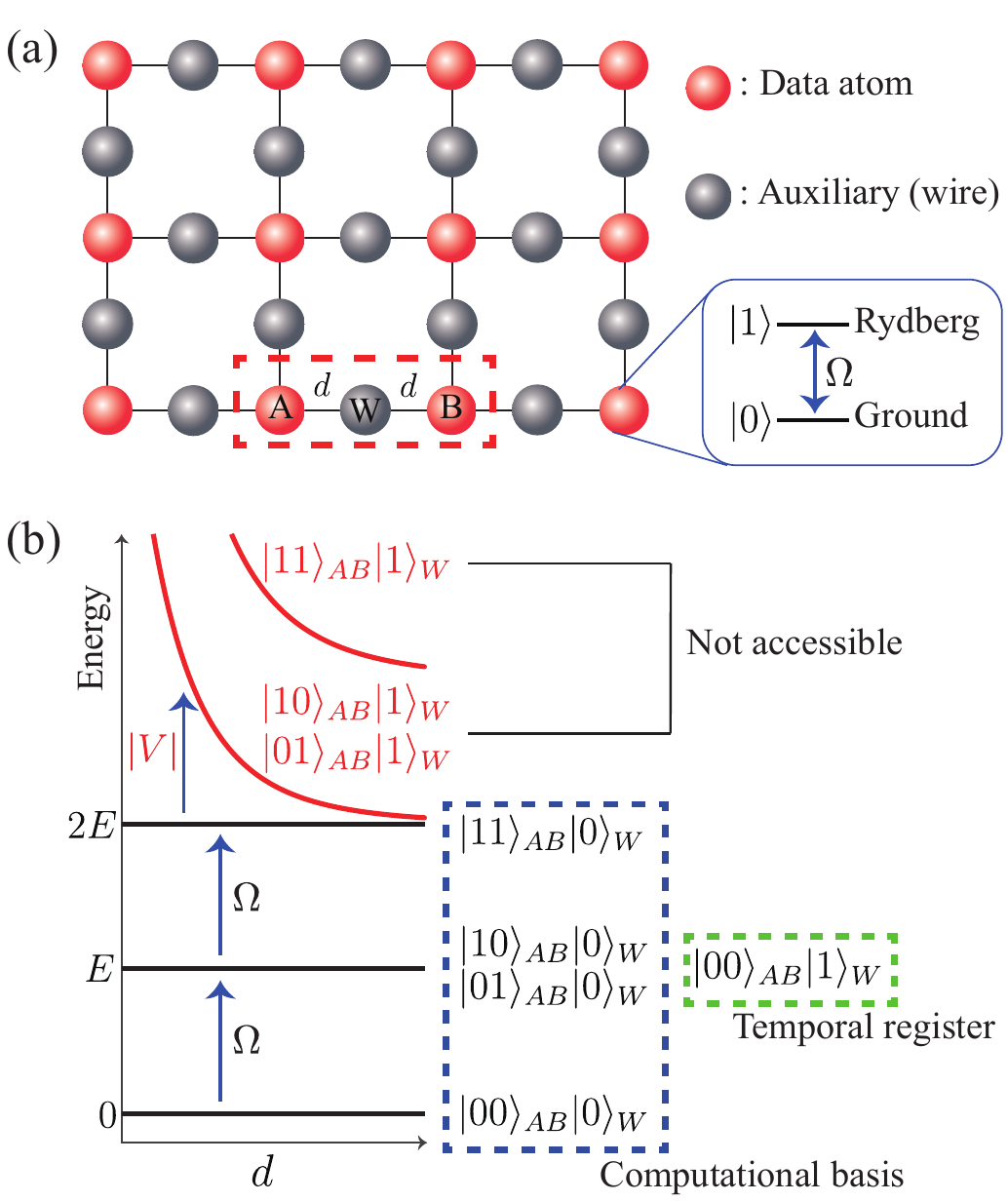}
\caption{The Rydberg wire gate scheme: (a) A 2D atomic array consists of data atoms (red spheres) and auxiliary (wire) atoms (grey spheres). Atomic ground state $\ket{0}$ and Rydberg state $\ket{1}$ are used for the two-level system of each atom. Wire atoms, e.g., $W$, mediate the couplings between two adjacent data atoms, e.g., $A$ and $B$ which are separated from $W$ by a distance $d$. (b) The energy level diagram of the three atoms, $A$, $W$, and $B$. We use four computational basis states, $\ket{00}_{\rm AB}\ket{0}_{\rm W}$, $\ket{01}_{\rm AB}\ket{0}_{\rm W}$, $\ket{10}_{\rm AB}\ket{0}_{\rm W}$, and $\ket{11}_{\rm AB}\ket{0}_{\rm W}$ (in the blue dashed rectangle), out of five accessible states including $\ket{00}_{\rm AB}\ket{1}_{\rm W}$ which is considered as a temporal register (in the light green dashed rectangle). The other states, $\ket{10}_{AB}\ket{1}_{W}$, $\ket{01}_{AB}\ket{1}_{W}$, and $\ket{11}_{AB}\ket{1}_{W}$ are not accessible due to the Rydberg blockade.}
\label{Fig1}
\end{figure}

\noindent
We aim to construct quantum gates with a sequence of individual-atom addressings in an array of atoms. We consider a two-dimensional (2D) array of atoms as shown in Fig.~\ref{Fig1}(a). In the Rydberg blockade regime, adjacent two atoms are inhibited from being excited to an anti-blockade state, $\ket{11}$, so the computational space of the two atoms is limited to $\{\ket{00},\ket{01},\ket{10}\}$ excluding $\ket{11}$, when the two-level system, $\{\ket{0},\ket{1}\}$, is defined with the ground and Rydberg states of each atom. However, because $\ket{11}$ is necessary for general quantum computation, we use auxiliary atoms (which we refer to as wire atoms, hereafter) to mediate couplings among the data atoms, which are illustrated with red (data atom) and grey (wire atom) spheres in Fig.~\ref{Fig1}(a).

In the three-atom system, $AWB$ in Fig.~\ref{Fig1}(a), $A$ and $B$ are data atoms and $W$ is the wire atom to couple $A$ and $B$. When the wire atom is excited to $\ket{1}$ only for data processing of $\ket{AB}$ and otherwise left to be $\ket{0}_{W}$, there are five computational base states $\ket{00}_{AB}\ket{0}_{W}$, $\ket{01}_{AB}\ket{0}_{W}$, $\ket{10}_{AB}\ket{0}_{W}$, $\ket{11}_{AB}\ket{0}_{W}$, and $\ket{00}_{AB}\ket{1}_{W}$. Here the first four base states are the computational basis for the two-data ($AB$) system and the last $\ket{00}_{AB}\ket{1}_{W}$ can be considered as a temporal register. There are three available atom-addressings:
\begin{subequations} \label{WAB}
\begin{eqnarray} 
\tilde{W}(\Theta,\phi)&=&e^{-\frac{i}{\hbar}\int H_{W} dt} ,\\
\tilde{A}(\Theta,\phi)&=&e^{-\frac{i}{\hbar}\int  \left( \frac{\hbar \Omega}{2}  \hat{n}_\phi \cdot \vec{\sigma}^{A} + V n_{W} n_{A}\right) dt},\\
\tilde{B}(\Theta,\phi)&=&e^{-\frac{i}{\hbar}\int  \left( \frac{\hbar \Omega}{2}  \hat{n}_\phi \cdot \vec{\sigma}^{B} + V n_{W} n_{B}\right) dt},
\end{eqnarray}
\end{subequations}
where $\Theta$ and $\phi$ are the Rabi rotation angle and axis, respectively. $H_W$ is the Hamiltonian of single-addressing of $W$ given by
\begin{equation} \label{HW}
{H}_{W} = \frac{\hbar \Omega}{2}  \hat{n}_\phi \cdot \vec{\sigma}^{W} + V n_{W} (n_{A}+n_{B}),
\end{equation}
in the Rydberg blockade regime of adjacent atoms, i.e., $d<d_B<\sqrt{2}d$, where $d$ and $d_B$ are the inter-atom and blockade distances, respectively. $\Omega$ is the Rabi frequency, $\hat n_\phi$ is the rotational axis defined by laser phase $\phi$, $V=C_6/d^6$ is the van der Waals interaction with coefficient $C_6$, and $\vec{\sigma}=(\sigma_x,\sigma_y,\sigma_z)$ is the Pauli vector and $n=(1-\sigma_z)/2$ is the excitation number.

It is noted that the atom-addressing operations in Eq.~\eqref{WAB} are three-qubit gates. We intend to use them for general quantum computations of the data $AB$ atoms. $\tilde{W}$ changes $\ket{00}_{AB}\ket{0}_{W}$ to $\ket{00}_{AB}\ket{1}_{W}$ and preserves all other states and their superpositions. So, the $\tilde{W}$ operation is the inverted controlled rotation gate, where $AB$ are control qubits and $W$ is the target qubit. The other three operators are reduced to single and two-atom rotations in the data-qubit ($AB$) basis as
\begin{subequations}
\begin{eqnarray}
{\bf R}_{A}\otimes {\bf I}_{B} &=& \bra{0}_{W} \tilde{A} \ket{0}_{W} , \\
{\bf I}_{A} \otimes {\bf R}_{B} &=& \bra{0}_{W} \tilde{B} \ket{0}_{W} ,\\
{\bf R}_{A} \otimes {\bf R}_{B} &=& \bra{0}_{W} \tilde{A}\tilde{B} \ket{0}_{W},
\end{eqnarray}\label{wab}
\end{subequations}
where ${\bf R}$ is the single-qubit rotation and ${\bf I}$ is the identity. 

\section{Standard one-qubit gates}\label{one-qubit} \noindent
With the atom-addressing operations, $\tilde{W}$, $\tilde{A}$, and $\tilde{B}$, in Eq.~\eqref{WAB}, we construct standard one-qubit gates which include Pauli gates, ${\bf X}$, ${\bf Y}$, and ${\bf Z}$, general rotation ${\bf R}(\Theta,\phi)$, Hadamard gate ${\bf H}$, and phase gate, ${\bf P}$. 

Pauli gates rotate the quantum state of one atom, while leaving other atoms unchanged. For the data atoms, $A$ and $B$, Pauli ${\bf X}$ gates are given by
\begin{subequations}
\begin{eqnarray}
{\bf X}_{A}\otimes {\bf I}_{B} &=& e^{i\alpha} \bra{0}_{W} \tilde X_{A} \ket{0}_{W}, \\
{\bf I}_{A} \otimes {\bf X}_{B} &=&  e^{i\alpha}  \bra{0}_{W} \tilde X_{B} \ket{0}_{W},
\end{eqnarray}
\end{subequations}
where $\tilde{X}_{A}=\tilde{A}(\pi,0)$, $\tilde{X}_{B}=\tilde{B}(\pi,0)$, and $\alpha=\pi/2$ is the global phase. Likewise, Pauli ${\bf Y}$ and ${\bf Z}$ gates are given by
\begin{subequations}
\begin{eqnarray}
  {\bf Y}_{A}\otimes {\bf I}_{B} &=& e^{i\alpha} \bra{0}_{W} \tilde Y_{A} \ket{0}_{W} ,\\
  {\bf I}_{A}  \otimes {\bf Y}_{B} &=& e^{i\alpha} \bra{0}_{W} \tilde Y_{B} \ket{0}_{W} ,\\
  {\bf Z}_{A} \otimes {\bf I}_{B} &=& e^{i\alpha} \bra{0}_{W} \tilde X_{A} \tilde Y_{A} \ket{0}_{W} ,\\
  {\bf I}_{A} \otimes {\bf Z}_{B} &=&  e^{i\alpha} \bra{0}_{W} \tilde X_{B} \tilde Y_{B} \ket{0}_{W},
\end{eqnarray}
\end{subequations}
where $\tilde{Y}_{\rm A}=\tilde{A}(\pi,\pi/2)$ and $\tilde{Y}_{\rm B}=\tilde{B}(\pi,\pi/2)$. 
General rotations are given by
\begin{subequations}
\begin{eqnarray}
{\bf R}_{A}(\Theta,\phi) \otimes {\bf I}_B &=& \bra{0}_W \tilde A(\Theta,\phi) \ket{0}_W, \\
{\bf I}_A \otimes {\bf R}_{B}(\Theta,\phi) &=& \bra{0}_W \tilde B(\Theta,\phi) \ket{0}_W.
\end{eqnarray}
\end{subequations}

Hadamard gate, ${\bf H}$, converts the quantum states, $\ket{0}$ and $\ket{1}$, to the superposition states, $\ket{+}=(\ket{0}+\ket{1})/\sqrt{2}$ or $\ket{-}=(\ket{0}-\ket{1})/\sqrt{2}$, respectively. The Hadamard gate is identical to $e^{i\pi/4}{\bf X}\sqrt{\bf Y}$, given by
\begin{subequations}
\begin{eqnarray}
{\bf H}_{A} \otimes {\bf I}_{B} &=&  e^{i\alpha} \bra{0}_{W} \tilde X_{A} \sqrt{\tilde Y_{A}} \ket{0}_{W}, \\
{\bf I}_{A} \otimes {\bf H}_{B} &=&  e^{i\alpha} \bra{0}_{W} \tilde X_{B} \sqrt{\tilde Y_{B}} \ket{0}_{W},
\end{eqnarray}
\end{subequations}
where $\sqrt{\tilde Y_{A}}=\tilde{A}(\pi/2,\pi/2)$ and $\sqrt{\tilde Y_{B}}=\tilde{B}(\pi/2,\pi/2)$ are peudo-Hadamard gates.
$\alpha=\pi/2$.

Phase gates, ${\bf P}_{A}(\phi)$ and ${\bf P}_{B}(\phi)$, are given by 
\begin{subequations}
\begin{eqnarray}
{\bf P}_{A}(\phi) \otimes {\bf I}_{B} &=& e^{i\phi/2} \bra{0}_{W} \tilde X_{A}^{\dagger} \tilde{A}(\pi,\phi/2) \ket{0}_{W}, \\
{\bf I}_{A} \otimes {\bf P}_{B}(\phi) &=& e^{i\phi/2} \bra{0}_{W} \tilde X_{B}^{\dagger} \tilde{B}(\pi,\phi/2) \ket{0}_{W}.
\end{eqnarray}
\end{subequations}
${\bf S}$ and ${\bf T}$ gates are obtained as ${\bf S}_{A} =  {\bf P}_{A}(\pi/2)$, ${\bf S}_{B} =  {\bf P}_{B}(\pi/2)$, ${\bf T}_{A} =  {\bf P}_{A}(\pi/4)$, and ${\bf T}_{B} =  {\bf P}_{B}(\pi/4)$.

The global phase, $\alpha$, of the above gates can be eliminated with a global phase gate. One example is
\begin{eqnarray}
{\bf Ph}(\alpha)&=&\bra{0}_{W} \tilde Y_{B} \tilde X_{W}^{\dagger} \tilde{W}(\pi,\alpha) \tilde Y_{{AB}}^{\dagger} \tilde X_{W}^{\dagger} \tilde{W}(\pi,\alpha) \nonumber \\ &\times& \tilde Y_{B}^{\dagger} \tilde X_{W}^{\dagger} \tilde{W}(\pi,\alpha) \tilde Y_{{AB}} \tilde X_{W}^{\dagger} \tilde{W}(\pi,\alpha) \ket{0}_{W}\label{phaseAlpha},
\end{eqnarray}
which is a combination of four two-qubit phase rotations, $\ket{00}\rightarrow e^{i{\alpha}}\ket{00}$ which is performed by $\tilde X_W^{\dagger} \tilde{W}(\pi,\alpha)$,  $\ket{01}\rightarrow e^{i{\alpha}}\ket{01}$  by  $\tilde Y_{B} \tilde X_W^{\dagger} \tilde{W}(\pi,\alpha) \tilde Y_{B}^{\dagger}$,  $\ket{10}\rightarrow e^{i{\alpha}}\ket{10}$ by  $\tilde Y_{A}^{\dagger} \tilde X_W^{\dagger} \tilde{W}(\pi,\alpha) \tilde Y_{A}$, and $\ket{11}\rightarrow e^{i{\alpha}}\ket{11}$ by $\tilde Y_{AB}^{\dagger} \tilde X_W^{\dagger} \tilde X_W(\pi,\alpha) \tilde Y_{AB}$, where $\tilde{Y}_{AB}$ denotes $\tilde{Y}_{A} \tilde{Y}_{B}$. 

\section{Standard 2-qubit gates}\label{two-qubit} \noindent
Now we consider standard two-qubit gates including the controlled-NOT gate, ${\bf CNOT}$, the swap gate, ${\bf SWAP}$, and the controlled-phase gate, ${\bf CP}$. 

\begin{figure}[htb]
\centering
\includegraphics [width=0.48\textwidth]{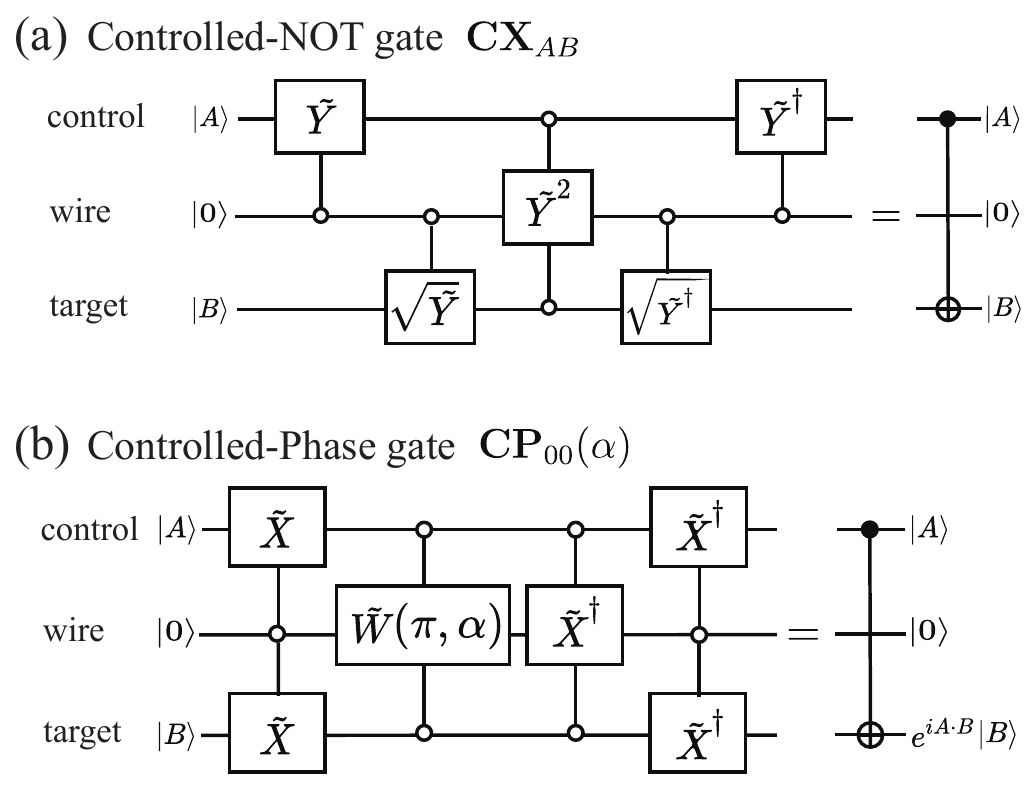}
\caption{Quantum circuits of (a) controlled-NOT gate, $\bold{CX}_{AB}$, and (b) controlled-phase gate, $\bold{CP}_{00}(\alpha)$}
\label{Fig2}
\end{figure}

Controlled-NOT gate, ${\bf CNOT}$, flips the target qubit (the second qubit) only when the control qubit (the first qubit) is in $\ket{1}$, i.e., $\ket{AB}\rightarrow \ket{{A}, {A}\oplus {B}}$, which is also the controlled ${\bf X}$ gate, i.e.,  ${\bf CNOT}={\bf CX}$. With atom-addressings, ${\bf CX}_{AB}$ and ${\bf CX}_{BA}$ are respectively given by
\begin{subequations}
\begin{eqnarray}
{\bf CX}_{AB} &=& \bra{0}_{W} \tilde Y_{A}^{\dagger} \sqrt{\tilde Y_{B}^{\dagger}} \tilde Y_{W}^{2} \sqrt{\tilde Y_{B}} \tilde Y_{A} \ket{0}_{W} , \\
{\bf CX}_{BA} &=& \bra{0}_{W} \tilde Y_{B}^{\dagger} \sqrt{\tilde Y_{A}^{\dagger}} \tilde Y_{W}^{2} \sqrt{\tilde Y_{A}} \tilde Y_{B} \ket{0}_{W},
\end{eqnarray} \label{CX}
\end{subequations}
of which the sequence can be understood as follows: In ${\bf CX}_{AB}$, $\tilde{Y}_{W}^2$ at the center works as an inverted-${\bf CZ}$ gate, which flips only the sign of the coefficient of $\ket{00}_{AB}\ket{0}_W$. When this is multiplied by $\tilde{Y}_{AB}$ from one side and by its hermitian conjugate from the other side, we get the controlled-$Z$ gate, i.e.,
\begin{equation}
{\bf CZ}_{AB} = {\bf CZ}_{BA} = \bra{0}_{W} \tilde Y_{AB}^{\dagger} \tilde Y_{W}^{2} \tilde Y_{AB} \ket{0}_{W}, \label{CZ} 
\end{equation}
which is then multiplied by $\sqrt{Y_{A}}$ and its hermitian conjugate, to attain ${\bf CX}_{\rm AB}$. The quantum circuit of ${\bf CX}_{\rm AB}$ is presented in Fig.~\ref{Fig2}(a).
Likewise, controlled-$Y$ gates are given by
\begin{subequations}
\begin{eqnarray}
{\bf CY}_{AB} &=& \bra{0}_{W} \tilde Y_{A}^{\dagger} \sqrt{\tilde X_{B}^{\dagger}} \tilde Y_{W}^{2} \sqrt{\tilde X_{B}} \tilde Y_{A} \ket{0}_{W}, \\
{\bf CY}_{BA} &=& \bra{0}_{W} \tilde Y_{B}^{\dagger} \sqrt{\tilde X_{A}^{\dagger}} \tilde Y_{W}^{2} \sqrt{\tilde X_{A}} \tilde Y_{B} \ket{0}_{W}.
\end{eqnarray}
\end{subequations}

${\bf SWAP}$ gate performs state swapping of two qubits, i.e., $\ket{AB} \rightarrow \ket{BA}$, which is also the exchange of the coefficients of $\ket{01}$ and $\ket{10}$. In our atom-addressing scheme, an ${\bf X}$-gate version of ${\bf SWAP}$ gate is given by
\begin{equation}
{\bf SWAP} = \bra{0}_W \tilde X_{A} \tilde X_{W} \tilde X_{AB} \tilde X_{W} \tilde X_{AB}^{\dagger} \tilde X_W \tilde X_{A}^{\dagger} \ket{0}_W,
\end{equation}
in which the first three-pulse combination, $X_{A}^{\dagger} \tilde X_W \tilde X_{A}^{\dagger}$, exchanges the coefficients of $\ket{10}_{AB}\ket{0}_W$ and $\ket{00}_{AB}\ket{1}_W$. The coefficient of $\ket{00}_{AB}\ket{1}_W$ is then exchanged with that of $\ket{01}_{AB}\ket{0}_W$ by the second combination, $\tilde X_{B}\tilde X_W\tilde X_B$, before the coefficient of $\ket{00}_{AB}\ket{1}_W$ is returned to $\ket{10}_{AB}\ket{0}_W$ by $\tilde X_{A}\tilde X_W\tilde X_A$.

Controlled-phase gate, ${\bf CP}(\alpha)$, puts the local phase of $\ket{11}$ of $AB$ data qubits. In our atom-addressing scheme, $W$-atom addressing, $\tilde{W}(\pi, \alpha)$, converts $\ket{00}_{AB}\ket{0}_{W}$ to $-i e^{i\alpha}\ket{00}_{AB}\ket{1}_{W}$ and $\tilde{W}(\pi, \pi) \tilde{W}(\pi, \alpha)$ converts $\ket{00}_{AB}$ to $e^{i\alpha}\ket{00}_{AB}$, so ${\bf CP}_{00}(\alpha)$, which puts the local phase of $\ket{00}$, is given by
\begin{equation}
{\bf CP}_{00}(\alpha) = \bra{0}_{W} \tilde X_{W}^{\dagger} \tilde{W}(\pi, \alpha) \ket{0}_{W}.
\label{CP}
\end{equation}
The quantum circuit of ${\bf CP}_{00}(\alpha)$ is presented in Fig.~\ref{Fig2}(b).
The standard ${\bf CP}(\alpha)={\bf CP}_{11}(\alpha)$ is therefore obtained by
\begin{eqnarray}
{\bf CP}(\phi) &=& \bra{0}_{W} \tilde X_{AB}^{\dagger} \tilde X_{W}^{\dagger} \tilde W(\pi,\phi) \tilde X_{AB} \ket{0}_{W},
\end{eqnarray}
where the ${\bf CP}_{00}(\alpha)$ in the middle is multiplied by $\tilde Y_{AB}$ from one side and by the conjugate of $\tilde Y_{AB}$ from the other side, which respectively exchanges and exchanges back the coefficients of $\ket{00}$ and $\ket{11}$. As a result, we get $\ket{11}\rightarrow e^{i\alpha}\ket{11}$. 
Similarly, ${\bf CP}_{01}(\phi)$ and ${\bf CP}_{10}(\phi)$ are  obtained as
\begin{eqnarray}
{\bf CP}_{01}(\phi) &=& \bra{0}_{\rm W} \tilde X_{\rm B}^{\dagger} \tilde X_{\rm W}^{\dagger} \tilde W(\pi,\phi) \tilde X_{\rm B} \ket{0}_{\rm W}, \\
{\bf CP}_{10}(\phi) &=& \bra{0}_{\rm W} \tilde X_{\rm A}^{\dagger} \tilde X_{\rm W}^{\dagger} \tilde W(\pi,\phi) \tilde X_{\rm A} \ket{0}_{\rm W}.
\end{eqnarray}

\section{Arbitrary two-qubit state generation} \label{arbitrary} \noindent
General two-qubit state generation is to find a unitary operation which transforms the initial state $\ket{00}_{AB}$ to an arbitrary two-qubit state, i.e.
\begin{equation}
U \ket{00}= a_{0} \ket{00} + a_{1} \ket{01} + a_{2} \ket{10} + a_{3} \ket{11}.
\end{equation}
The above $U$ can be in principle constructed with single- and two-qubit gates. Also, it is sufficient to define general rotations and at least one inversion operation among the two-qubit base states, $\{\ket{00},\ket{01},\ket{10},\ket{11}\}$ of $AB$ atoms.

Inversion operations are the reflection of the two-qubit state vector about a given plane. For example, ${\bf CZ}$ inverts the state vector about the plane orthogonal to $\ket{11}$, i.e., ${\tilde M}_{11}={\bf CZ}$. Likewise, ${\tilde M}_{00}={\bf CP}_{00}(\pi)$, ${\tilde M}_{01}={\bf CP}_{01}(\pi)$, and ${\tilde M}_{10}={\bf CP}_{{10}}(\pi)$.

General rotations are the base-pair rotation between a pair of two-qubit base states, i.e., $\tilde R_{jk} (\Theta,\phi)\ket{j}=\cos\frac{\Theta}{2}\ket{j} -i e^{i\phi} \sin\frac{\Theta}{2} \ket{k}$ for $j,k\in\{\ket{00},\ket{01},\ket{10},\ket{11}\}$. $\tilde R_{00,01}(\Theta,\phi)$ rotates the quantum information stored in the base pair, $\ket{00}$ and $\ket{01}$, which is for example given by
\begin{equation}
\tilde R_{00,01}(\Theta,\phi) = \bra{0}_{W} \tilde{X}_{W} \tilde X_{B} \tilde W(\Theta,-\phi) \tilde X_{B}^\dagger \tilde X_{W}^\dagger \ket{0}_{W}, \label{R01} 
\end{equation}
where the first two $\pi$-pulse operations, $\tilde X_B^\dagger$ and $\tilde X_W^\dagger$, perform $\ket{00}_{AB}\ket{0}_W \rightarrow \ket{00}_{AB}\ket{1}_W$ and $\ket{01}_{AB}\ket{0}_W \rightarrow \ket{00}_{AB}\ket{0}_W$, respectively, which means that the quantum state of $B$ atom is transferred to $W$ atom. Then the state vector of $W$ atom is rotated by $\tilde W(\Theta,-\phi)$ and transferred back to $B$ atom by the last two $\pi$-pulse operations. Similarly, other rotations can be obtained as follows: 
\begin{widetext}
\begin{subequations}
\begin{eqnarray}
\tilde R_{00,11}(\Theta,\phi) &=& \bra{0}_W \tilde{X}_W \tilde{X}_{AB} \tilde{W}(\Theta,-(\phi+\pi/2)) \tilde{X}_{AB}^\dagger \tilde{X}_W^\dagger \ket{0}_W,  \\
\tilde R_{01,10}(\Theta,\phi) &=& \bra{0}_W \tilde{X}_B \tilde{X}_W \tilde{X}_{AB} \tilde{W}(\Theta,-(\phi+\pi/2))  \tilde{X}_{AB}^\dagger  \tilde{X}_W^\dagger \tilde{X}_B^\dagger \ket{0}_W , \label{R12}, \\
\tilde R_{01,11}(\Theta,\phi) &=& \bra{0}_W \tilde{X}_{B} \tilde{X}_W \tilde{X}_A 
\tilde{W}(\Theta,-\phi) \tilde{X}_A^\dagger \tilde{X}_W^\dagger \tilde{X}_{B}^\dagger \ket{0}_W, \\
\tilde R_{10,11}(\Theta,\phi) &=& \bra{0}_W \tilde{X}_A \tilde{X}_W \tilde{X}_B,
\tilde{W}(\Theta,-\phi) \tilde{X}_B^\dagger \tilde{X}_W^\dagger \tilde{X}_A^\dagger \ket{0}_W. 
\end{eqnarray}
\end{subequations}
\end{widetext} 

\section{Multi-qubit gates} \label{multi} \noindent
While multi-qubit gates can be decomposed to a sequence of single- and two-qubit elementary gates, standard three-qubit gates require many elementary gates; for example, a Toffoli gate needs 15 or 17 elementary gates. In the following, we consider the possibilities of using wire-atom arrangements which can reduce the number of gates significantly for Toffoli and ${\bf CCZ}$ gates. 

\begin{figure}[t]
\centering
\includegraphics [width=0.48\textwidth]{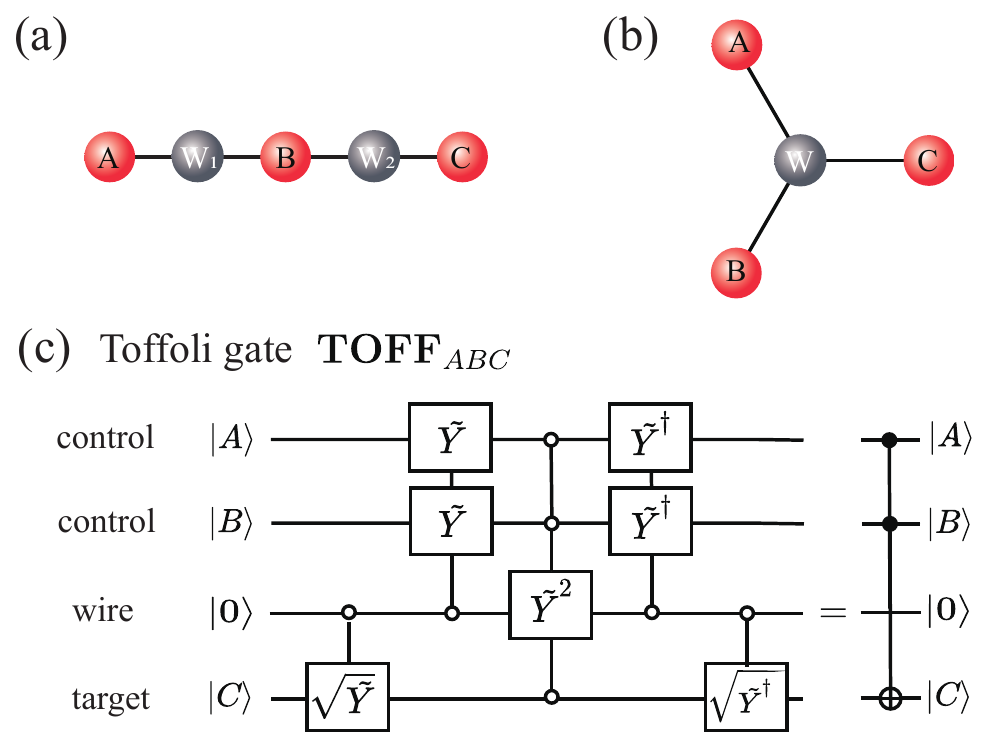}
\caption{(a) An 5-atom chain and (b) an Y-shape atomic array to implement multi-qubit wire gates. (c) Quantum circuit of the Toffoli gate $\bold{TOFF}_{ABC}$ for the control atoms $A$, $B$ and the target atom $C$. }
\label{Fig3}
\end{figure}

If we use the simple linear configuration, as in Fig.~\ref{Fig3}(a), of $ABC$ data atoms and two wire atoms $W_1$ and $W_2$, their pulse-sequence solutions, e.g., for Toffoli and ${\bf CCZ}$ gates, are rather complicated:
\begin{eqnarray}
{\bf CCZ} &=& \bra{00}_{{W}_{12}} \sqrt{\tilde{Y}_{C}} \tilde{Y}_{AB} \nonumber \tilde{X}_{W_{2}}^{\dagger} \tilde{Y}_{W_{1}} \tilde{X}_{BC}^{\dagger} \sqrt{\tilde{X}_{W_{2}}} \tilde{X}_{BC}^{2} \\ &\times& \sqrt{\tilde{X}_{W_{2}}^{\dagger}} \tilde{X}_{BC}^{\dagger}   \tilde{X}_{W_{12}} \tilde{Y}_{AB}^{\dagger} \sqrt{\tilde{Y}_{C}^{\dagger}} \ket{00}_{{W}_{12}}, \\
{\bf TOFF} &=& \bra{00}_{{W}_{12}} \sqrt{\tilde{Y}_{B}^{\dagger} \tilde{Y}_{C}} \tilde{Y}_{AB} \tilde{X}_{W_{2}}^{\dagger} \tilde{Y}_{W_{1}} \tilde{X}_{BC}^{\dagger} \sqrt{\tilde{X}_{W_{2}}}\tilde{X}_{BC}^{2} \nonumber \\ &\times&   \sqrt{\tilde{X}_{W_{2}}^{\dagger}} \tilde{X}_{BC}^{\dagger} \tilde{X}_{W_{12}} \tilde{Y}_{AB}^{\dagger} \sqrt{\tilde{Y}_{C}^{\dagger} \tilde{Y}_{B}} \ket{00}_{{W}_{12}}.
\end{eqnarray}

Instead, if we use the $Y$-shape configuration, as shown in Fig.~\ref{Fig3}(b), which has one wire-atom, $W$, which couples the all three data atoms, $ABC$, simultaneously, their solutions are simple, given as the extensions of ${\bf CX}$ and ${\bf CZ}$ in Eqs.~\eqref{CX} and \eqref{CZ}. The ${\bf CCZ}$ utilizes the fact that $\bra{0}_{W}\tilde{Y}_{W}^2 \ket{0}_{W}$ is the inverted-${\bf CCZ}$, to attain
\begin{equation}
{\bf CCZ} = \bra{0}_{W} \tilde Y_{ABC}^\dagger \tilde Y_{W}^2 \tilde Y_{ABC} \ket{0}_{W}, \label{CCZ}
\end{equation}
where $\tilde Y_{ABC}=\tilde Y_{A}\tilde Y_{B}\tilde Y_{C}$ and  $\tilde Y_{ABC}^\dagger$ are for the bitwise flip and flip-back of the data atoms, applied before and after to change the inverted-${\bf CCZ}$ to ${\bf CCZ}$. The Toffoli gate of $AB$ controls and $C$ target is also obtained as
\begin{equation}
{\bf TOFF}_{\rm ABC} = \bra{0}_{\rm W} \sqrt{\tilde Y_{\rm C}^{\dagger}} \tilde Y_{\rm AB}^{\dagger}  \tilde Y_{\rm W}^2  \tilde Y_{\rm AB}\sqrt{\tilde Y_{\rm C}} \ket{0}_{\rm W},
\end{equation}
where $\sqrt{\tilde Y_{\rm C}^{\dagger}}$ and $\sqrt{\tilde Y_{\rm C}}$ on the both ends are the pseudo-Hadamard and its inverse acting on the target. The quantum circuit of ${\bf TOFF}_{\rm ABC}$ is presented in Fig.~\ref{Fig3}(c).

\section{Discussions and conclusion}\label{discussion} \noindent 
{\it Experimental implementation}: Rydberg wire gates introduced above can be implemented in optical-tweezer atomic systems, which have been previously demonstrated elsewhere~\cite{Jo2020,Levine2019,Graham2019}. As an example, we consider three rubidium ($^{87}$Rb) atoms arranged in the linear chain geometry. Once the single atoms are loaded to individual tweezers from magneto-optical-trap, the atoms are prepared to one of magnetic sublevels in hyperfine ground states as the ground state $\ket{0}$ (for example, $\ket{0} = \ket{5S_{1/2}, F=2, m_F=2}$). The states $\ket{0}$ and $\ket{1}$ are coupled by Rydberg state excitation lasers, and in general two-photon excitation is used to transit to $\ket{nS}$ or $\ket{nD}$ Rydberg levels via $\ket{5P_{3/2}}$ with 780~nm and 480~nm lights. For $\ket{1} = \ket{69S_{1/2}}$ the atoms undergo van der Waals interaction, and the interaction strength when the interatomic distance $d=7~\mu$m becomes $V= \left |C_6 \right | / d^6 = (2\pi)6.2$~MHz, where $C_6 = -(2\pi)732$~GHz. Individual atom-addressings to couple between $\ket{0}$ and $\ket{1}$ can be implemented by diffracting multiple laser beams from an acousto-optic modulator (AOM), then focusing to individual atoms. The switching of individual beams can be done by controlling amplitude and frequency of radio-frequency wave to AOM. The individual addressing lasers can be either ground-Rydberg resonant lasers~\cite{Graham2019} or far-detuned lasers~\cite{Omran2019}, in which the latter suppress the Rydberg state excitation with additional AC Stark shift combined with global resonant lasers.

\begin{figure}[h]
\centering
\includegraphics [width=0.5\textwidth]{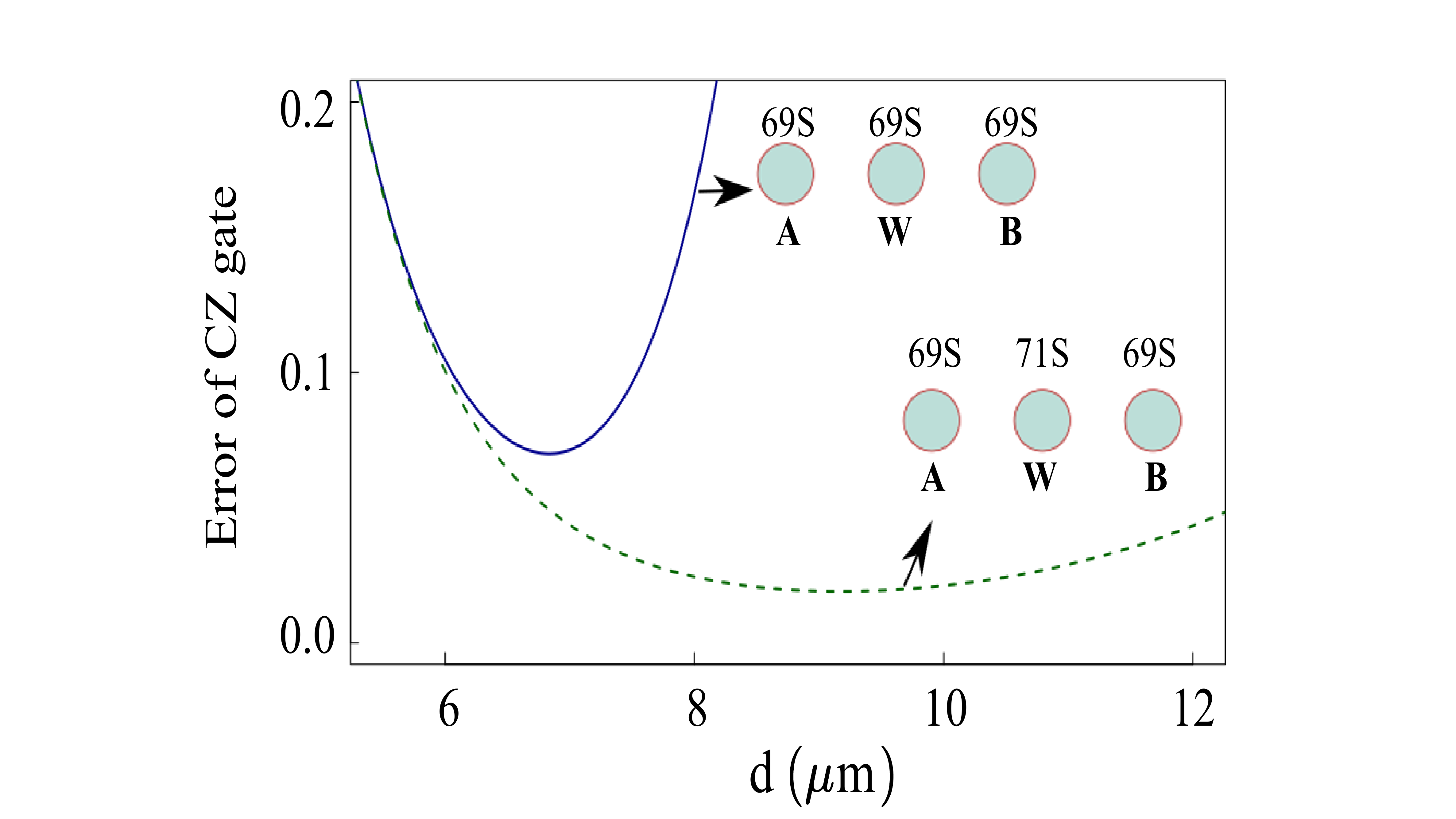}
\caption{Performance estimation of the $\bold{CZ}$ gate for the present van der Waals scheme (solid line) in comparison with the F\"{o}rster resonance scheme (dashed line)}
\label{Fig4}
\end{figure}

\noindent {\it Gate performance}: The performance of the Rydberg wire gate schemes can be estimated with numerical calculations. In Fig.~\ref{Fig4}, we estimate the average fidelity of ${\bf CP}_{00}(\pi)$ gate for all initial states $\left\{ \ket{00}_{AB}, \ket{01}_{AB}, \ket{10}_{AB}, \ket{11}_{AB} \right\}$ using time-dependent Schr\"{o}dinger equations. For $\ket{1}=\ket{69S_{1/2}, m_j=1/2}$, the results with respect to the interatomic distance are shown with the solid line in Fig. \ref{Fig4}. For Rabi frequency $\Omega=(2\pi) 2$~MHz, the gate duration is 0.5~$\mu$s. It is expected that the maximum fidelity $\mathcal{F}$ can be reached to 94~$\%$ when the lattice constant is around 6.8~$\mu$m.

\noindent {\it Gate imperfection sources}: The sources of finite infidelities related to the Rydberg atomic properties can be characterized. The finite lifetime of Rydberg state gives imperfection to the transition to $\ket{1}$. For the lifetime of $\ket{1}$ to be $\tau$, this gives the average gate error $\frac{9\pi}{4 \Omega \tau}$~\cite{Shi2017}. Another source of the gate infidelity is the Rydberg blockade error: as the Rydberg interaction strength is proportional to $1/d^6$, the interaction strength within the blockade distance $d_B$ is finite, and there is non-zero residual interactions outside. For the interaction strength,  $V$, between a nearest neighbor Rydberg atomic pair, the gate error is given by $\frac{\hbar^2\Omega^2}{2V^2}$ for the initial state $\ket{10}_{AB}, \ket{01}_{AB}$ and $\frac{\hbar^2\Omega^2}{8V^2}$ for $\ket{11}_{AB}$~\cite{Saffman2005,Shi2021qst}. In addition, the phase shift $\frac{2\pi V_2}{\hbar \Omega}$ occurs for the initial state $\ket{11}_{AB}$, due to the nonzero interactions between atom $A$ and $B$. Considering all these error budgets, we estimate the average fidelity error as
\begin{equation} \label{error}
1-\mathcal{F}= \frac{9\pi}{4\Omega\tau} +  \frac{9\hbar^2\Omega^2}{32V^2} +\frac{\pi V}{128\hbar\Omega},
\end{equation}
where the terms denote the Rydberg state decay error, the Rydberg blockade error, and the residual interaction error, respectively. Their estimated infidelity contributions are $4 \times 10^{-3}$, $2.04 \times 10^{-2}$ and $9.12 \times 10^{-2}$, respectively, at $d=6.8~\mu$m. 

\noindent {\it Toward higher fidelity gates}: We discuss methods to improve the gate fidelity to suppress the last two errors in Eq.~\eqref{error}.
One approach is to utilize the dipole-dipole interaction by F\"{o}rster resonance between a nearest neighbour atomic pair. Near the principal quantum number $n=69$ discussed above, there exist two transition channels between Rydberg pair states, $\ket{69S_{1/2} + 71S_{1/2}} \leftrightarrow \ket{69P_{3/2} + 70P_{1/2}}$ and $\ket{69S_{1/2} + 71S_{1/2}} \leftrightarrow \ket{69P_{1/2} + 70P_{3/2}}$ by the dipole-dipole interaction, with F\"{o}rster defects of 6.6 and 19.7~MHz, respectively~\cite{Weber2017}. This induces the dipole-dipole interaction with the strength of $V' = C_3/d^3$, where $C_3 = (2\pi) 12.32$~GHz.$\mu\textup{m}^3$, with the interatomic distance less than the crossover distance 11~$\mu$m~\cite{Walker2008}. In realizing the ${\bf CP}_{00}(\pi)$ gate, the atom $W$ is to be  excited to $\ket{1'}=\ket{71S_{1/2}, m_j=1/2}$ state, while the data atoms $A$ and $B$ are excited to $\ket{1}$. Then the interaction strength between $A(B)$ and $W$ is increased due to the F\"{o}rster resonance, so the interatomic distance can also be increased. This further reduces the long range residual van der Waals interaction between $A$ and $B$, thus the gate infidelity can be suppressed. In Fig.~\ref{Fig4}, we illustrate the imporved performance of the ${\bf CP}_{00}(\pi)$ gate of the dipole-dipole interaction (the dashed line). The overall fidelities $\mathcal{F}$ are increased compared to the previous example, and the maximum is reached to 98~$\%$ at $d=9.17~\mu$m.

\noindent {\it Weakness of the Rydberg wire gates}: The weakness of the present scheme is that the Rydberg states are not stable. There is a constant decay process occurring during the quantum control process. However, for a fast quantum control process, the decay-induced error can be relatively small for the decay error is proportional to the Rydberg superposition time. Moreover, quantum error correction can in principle be executed by the gates shown in this paper, so that the error during the control process can be corrected. Because both the main control process and error correction are fast thanks to the fast pulsed operations of quantum wire gates, the overall speed to reach a wanted computational result can still surpass the traditional method of coding information with stable hyperfine-Zeeman sub-states. 

In summary, Rydberg wire gates are proposed, which utilize auxiliary atoms to couple data atoms. By coding information with a ground-state qubit state and a Rydberg qubit state, the universal gate set can be realized based on strong, local interactions of neutral Rydberg atoms. The gates are realized by fast laser excitation of Rydberg states, so that their speed can be fast, and well-separated data atoms can be rapidly entangled. Fast entangling operations are important basic elements in a quantum circuit for large-scale quantum computation, and long-distance entanglement can greatly simply complex operations between distant qubits in the array. The new idea of Rydberg wire gates can bring new prospective in neutral-atom quantum science and technology. 

\begin{acknowledgements}
This research was supported by Samsung Science and Technology Foundation (Grant No. SSTF-BA1301-52), National Research Foundation of Korea (Grant No. 2017R1E1A1A0107430), and Natural Science Foundation of China (Grant No. 12074300).
\end{acknowledgements}

\end{document}